\RequirePackage{ifpdf}
\ifpdf 
\documentclass[pdftex]{sigma}
\else
\documentclass{sigma}
\fi

\def\sn{\mathrm{sn\,}}
\def\cn{\mathrm{cn\,}}

\def\dn{\mathrm{dn\,}}

\usepackage{eucal}

\def\sn{\mathrm{sn\,}}
\def\cn{\mathrm{cn\,}}

\def\dn{\mathrm{dn\,}}

\def\openone{\leavevmode\hbox{\small1\kern-3.3pt\normalsize1}}
\def\sech{\mbox{sech\,}}

\def\openone{\leavevmode\hbox{\small1\kern-3.3pt\normalsize1}}

\allowdisplaybreaks

\usepackage{epsf}

\def\newpic#1{%
   \def\emline##1##2##3##4##5##6{%
      \put(##1,##2){\special{em:point #1##3}}%
      \put(##4,##5){\special{em:point #1##6}}%
      \special{em:line #1##3,#1##6}}}

\newpic{}

\def\openone{\leavevmode\hbox{\small1\kern-3.3pt\normalsize1}}
\def\sech{\mathop{\rm sech}\nolimits}

\begin{document}

\allowdisplaybreaks

\renewcommand{\PaperNumber}{071}

\FirstPageHeading

\numberwithin{equation}{section}

\ShortArticleName{Exact Solutions for Equations
of Bose--Fermi Mixtures}

\ArticleName{Exact Solutions for Equations
of Bose--Fermi Mixtures\\ in One-Dimensional Optical Lattice}

\Author{Nikolay A. KOSTOV~$^{\dag}$, Vladimir S. GERDJIKOV~$^\ddag$ and
Tihomir I. VALCHEV~$^\ddag$}

\AuthorNameForHeading{N.A. Kostov, V.S. Gerdjikov and T.I. Valchev}

\Address{$^\dag$~Institute of Electronics, Bulgarian Academy of
Sciences,\\
$\phantom{^\dag}$~72 Tsarigradsko chaussee, 1784 Sofia, Bulgaria}
\EmailD{\href{mailto:nakostov@inrne.bas.bg}{nakostov@inrne.bas.bg}}

\Address{$^\ddag$~Institute for Nuclear Research and Nuclear
Energy, Bulgarian Academy of Sciences,\\
$\phantom{^\dag}$~72 Tsarigradsko chaussee,
1784 Sofia, Bulgaria}

\EmailD{\href{mailto:gerjikov@inrne.bas.bg}{gerjikov@inrne.bas.bg}, \href{mailto:valtchev@inrne.bas.bg}{valtchev@inrne.bas.bg}}

\ArticleDates{Received March 30, 2007, in f\/inal form May
17, 2007; Published online May 30, 2007}

\Abstract{We present two new families of stationary solutions for
equations of Bose--Fermi mixtures with an elliptic function
potential with modulus $k$. We also discuss particular cases when
the quasiperiodic solutions become periodic ones. In the limit of
a sinusoidal potential ($k\to 0$) our solutions model a quasi-one
dimensional quantum degenerate Bose--Fermi mixture trapped in
optical lattice. In the limit $k\to 1$  the solutions are
expressed by hyperbolic function solutions (vector solitons). Thus
we are able to obtain in an unif\/ied way quasi-periodic and
periodic waves, and solitons. The precise conditions for existence of every class of solutions are derived. There are
indications that such waves and localized objects may be observed
in experiments with cold quantum degenerate gases.}


\Classification{37K20; 35Q51; 74J30; 78A60}

\section{Introduction}

Over the last decade, the f\/ield of cold degenerate gases has been
one of the most active areas in physics. The discovery of
Bose--Einstein Condensates (BEC) in 1995 (see e.g.\
\cite{dgps99,mo06}) greatly stimulated research of ultracold
dilute Boson-Fermion mixtures. This interest is driven by the
desire to understand strongly interacting and strongly correlated
systems, with applications in solid-state physics, nuclear
physics, astrophysics, quantum computing, and nanotechnologies.

An important property of Bose--Fermi mixtures wherein the fermion
component is dominant is that the mixture tends to exhibit
essentially three-dimensional character even in a strongly
elongated trap. During the last decade, great progress has been
achieved in the experimental realization of Bose--Fermi  mixtures
\cite{mrr02,sfc01}, in particular Bose--Fermi  mixtures in
one-dimensional lattices. Optical lattices provide a powerful tool
to manipulate matter waves, in particular solitons.  The Pauli
exclusion principle results in the extension of the fermion cloud
in the transverse direction over distances comparable to the
longitudinal dimension of the excitations. It has been shown
recently, however, that the quasi-one-dimensional situation can
nevertheless be realized in a Bose--Fermi mixture due to strong
localization of the bosonic component \cite{tw00,skk06}. With account of the ef\/fectiveness of the optical lattice in managing
systems of cold atoms, their ef\/fect on the dynamics of Bose--Fermi
mixtures is of obvious interest. Some of the aspects of this
problem have already been explored within the framework of the
mean-f\/ield approximation. In particular,  the dynamics of the
Bose--Fermi mixtures were explored from the point of view of
designing quantum dots \cite{s05}. The localized states of
Bose--Fermi mixtures with attractive (repulsive) Bose--Fermi
interactions are viewed as a matter-wave realization of quantum
dots and antidots. The case of Bose--Fermi mixtures in optical
lattices is investigated in detail and the existence of gap
solitons is shown. In particular, in \cite{s05} it is obtained
that the gap solitons can trap a number of fermionic bound-state
levels inside both for repulsive and attractive boson-boson
interactions.  The  time-dependent dynamical
mean-f\/ield-hydrodynamic model to study the formation of fermionic
bright solitons in a trapped degenerate Fermi gas mixed with
a~Bose--Einstein condensate in a quasi-one-dimensional cigar-shaped
geometry is proposed in~\cite{a05}. Similar model is used to study
mixing-demixing in a degenerate fermion-fermion mixture in~\cite{a06}.
Modulational instability, solitons and periodic waves
in a model of quantum degenerate boson-fermion mixtures are
obtained in~\cite{bpv07}.

Our aim is to derive two new classes of quasi-periodic exact
solutions of the time dependent mean f\/ield equations of Bose--Fermi
mixture in one-dimensional lattice. We also study some limiting
cases of these solutions. The paper is organized as follows. In
Section~\ref{sec:1} we give the basic equations.
Section~\ref{sec:2} is devoted to derivation of the f\/irst class
quasi-periodic solutions with non-trivial phases. A system of
$N_{f}+1$ equations, which reduce quasi-periodic solutions to
periodic are derived. In Section~\ref{sec:3} we present second
class (type B) nontrivial phase solutions.  In Section~\ref{sec:4}
we obtain 14 classes of elliptic solutions. Section~\ref{sec:5} is
devoted to two special limits, to hyperbolic and trigonometric
functions. In Section~\ref{sec:7} preliminary
results about the linear stability of solutions are given.
Section~\ref{sec:8} summarizes the main conclusions of the paper.

\section{Basic equations}\label{sec:1}

At mean f\/ield approximation we consider the following $N_{f} + 1$
coupled equations \cite{kb04,s05,bskk06,bpv07}
\begin{gather}
 i\hbar \frac{\partial \Psi^{b}}{\partial t}+\frac{1}{2 m_{\rm{B}}}
    \frac{\partial^2 \Psi^{b}}{\partial x^2}-V\Psi^{b}- g_{\rm{BB}}
    |\Psi^{b}|^2\Psi^{b}
    - g_{\rm{BF}}\rho_{f}\Psi^{b}=0, \label{bf1}\\\label{bf2}
  i\hbar \frac{\partial \Psi^{f}_j}{\partial t}+
    \frac{1}{2 m_{\rm{F}}}\frac{\partial^2 \Psi^{f}_j}{\partial
    x^2}-V\Psi^{f}_j- g_{\rm{BF}}|\Psi^{b}|^2\Psi^{f}_j=0,
    \end{gather}
where $\rho_{f}=\sum\limits_{i=1}^{N_{f}}|\Psi^{f}_{i}|^2$ and
\begin{gather*}
g_{\rm{BB}}=\frac{2 a_{\rm{BB}}}{a_{\rm{s}}},\qquad
g_{\rm{BF}}=\frac{2 a_{\rm{BF}}}{a_{\rm{s}}\alpha}, \qquad
\alpha=\frac{m_{\rm{B}}}{m_{\rm{F}}},\qquad
a_{\rm{s}}=\sqrt{\frac{\hbar}{m_{\rm{B}}\omega_{\bot}} },
\end{gather*}
$a_{\rm{BB}}$ and $a_{\rm{BF}}$ are the scattering lengths for
$s$-wave collisions for boson-boson and boson-fermion
interactions, respectively. In recent experiments
\cite{mi02,gi04} the quantum degenerate mixtures of
${}^{40}\rm{K}$ and ${}^{87}\rm{Rb}$ are studied where
$m_{\rm{B}}=87 m_{\rm{p}}$ , $m_{\rm{B}}=40 m_{\rm{p}}$ and
$\omega_{\bot}=215$ $\rm{Hz}$. Equations~(\ref{bf1}), (\ref{bf2}) have
been studied numerically in~\cite{kb04}. The formation of
localized structures containing bosons and fermions has been
reported in the particular case in which the interspecies
scattering length $a_{\rm{BF}}$ is negative, which is the case of
the ${}^{40}$K-${}^{87}$Rb mixture.
 An appropriate class of periodic potentials to model the
quasi-$1D$ conf\/inement produced by a standing light wave is given
by~\cite{bcdn01}
\begin{gather*}
V=V_{0}\sn^2(\alpha x,k) ,
\end{gather*}
where $\sn(\alpha x,k)$ denotes the Jacobian elliptic sine
function with elliptic modulus $0\leq k\leq 1$.

Experimental realization of two-component Bose--Einstein
condensates have stimulated considerable attention in general
\cite{mdfmm00} and in particular in the quasi-1D regime
\cite{ba01,Deco} when the Gross--Pitaevskii equations for two
interacting Bose--Einstein condensates reduce to coupled nonlinear
Schr\"{o}dinger (CNLS) equations with an external potential. In
specif\/ic cases the two component CNLS equations can be reduced to
the Manakov system~\cite{Manakov} with an external potential.

Important role in analyzing these ef\/fects was played by the
elliptic and periodic solutions of the above-mentioned equations.
Such solutions for the one-component nonlinear Schr\"{o}dinger
equation are well known, see \cite{book} and the numerous
references therein. Elliptic solutions for the CNLS and Manakov
system were derived in \cite{pp*99,ceek00,EilEnoKo}.

In the presence of external elliptic potential explicit stationary
solutions for NLS were derived in \cite{bcdn01,ckr01,bcdkp01}.
These results were generalized to the $n$-component CNLS in
\cite{Deco}. For $2$-component CNLS explicit stationary solutions
are derived in \cite{kegks04}.

\section{Stationary solutions with non-trivial
phases}\label{sec:2}

We restrict our attention to stationary solutions of these CNLS
\begin{gather}
\Psi^{b}(x,t)= q_{0}(x)\exp\left(-i\frac{\omega_{0}}{\hbar}t+
i\Theta_{0}(x)+i\kappa_{0} \right),\label{Abf1} \\  \Psi^{f}_j(x,t)=
q_{j}(x)\exp\left(-i\frac{\omega_{j}}{\hbar}t+ i\Theta_{j}(x)+i\kappa
_{0,j} \right), \label{Abf2}
\end{gather}
where $j=1,\ldots, N_{f}$, $\kappa _{0}$, $\kappa _{0,j} $, are
constant phases, $q_{j}$ and $\Theta_{0}$, $\Theta _j(x) $ are
real-valued functions connected by the relation
\begin{gather}
\Theta_{0}(x)={\mathcal{C}}_{0}\int_{0}^{x}
\frac{dx'}{q^2_{0}(x')}, \qquad
\Theta_{j}(x)={\mathcal{C}}_{j}\int_{0}^{x}
\frac{dx'}{q^2_{j}(x')}, \label{Theta}
\end{gather}
${\mathcal{C}}_{0},{\mathcal{C}}_{j}$, $j=1,\ldots, N_{f}$ being
constants of integration. Substituting the ansatz (\ref{Abf1}),
(\ref{Abf2}) in equations (\ref{bf1}) and separating the real and
imaginary part we get
\begin{gather}\label{sBFV1}
\frac{1}{2 m_{\rm{B}}} q_{0}^{3} q_{0xx} -g_{\rm{BB}} q_{0}^{6}-V
q_{0}^{4} - g_{\rm{BF}} \left(\sum_{i=1}^{N_{f}}q_{i}^2\right)
q_{0}^{4}+\omega_{0} q_{0}^{4}= \frac{1}{2
m_{\rm{B}}}{\mathcal{C}}_{0}^{2},
\\ 
\frac{1}{2 m_{\rm{F}}}q_{j}^{3}q_{jxx} - g_{\rm{BF}}q_{0}^2
q_{j}^{4} - V q_{j}^{4} +\omega_{j} q_{j}^{4}= \frac{1}{2
m_{\rm{F}}}{\mathcal{C}}_{j}^{2} .\nonumber
\end{gather}

We seek solutions for $q_{0}^{2}$ and $q_{j}^{2}$, $j=1,\ldots,
N_{f}$ as a quadratic function of $\sn(\alpha x,k)$:
\begin{gather}
q_{0}^{2}= A_{0}\sn^{2}(\alpha x,k)+B_{0}, \qquad  q_{j}^{2}=
A_{j}\sn^{2}(\alpha x,k)+B_{j}. \label{AbfV1}
\end{gather}
Inserting (\ref{AbfV1}) in (\ref{sBFV1}) and equating the
coef\/f\/icients of equal powers of $\sn(\alpha x,k)$ results in the
following relations among the solution parameters $\omega_{j}$, $
\mathcal{{C}}_{j}$, $A_{j}$ and $B_{j}$ and the characteristic of
the optical lattice $V_{0}$, $\alpha$ and $k$:
\begin{gather}\label{A}
A_{0}=\frac{\alpha^2 k^2- m_{F}V_{0}}{m_{\rm{F}} g_{\rm{BF}}},\qquad
\sum_{j=1}^{N_{f}} A_{j}=\frac{\alpha^2 k^2}{g_{\rm{BF}}}\left(\frac{1}{m_{\rm{B}}
}-\frac{g_{\rm{BB}}}{m_{\rm{F}} g_{\rm{BF}}}\right)-
\frac{V_{0}}{g_{\rm{BF}}}\left(1-\frac{g_{\rm{BB}}}{g_{\rm{BF}}}\right),
\\
\omega_{0}=\frac{\alpha^2 (k^2+1)}{2 m_{\rm{B}}}+g_{\rm{BB}}B_{0}+g_{\rm{BF}}\sum_{i=1}^{N_{f}}
B_{i} +\frac{\alpha^2 k^2}{2 m_{\rm{B}}}\frac{B_{0}}{A_{0}},\nonumber\\
\omega_{j}=\frac{\alpha^2 (k^2+1)}{2 m_{\rm{F}}}+ g_{\rm{BF}} B_{0} +\frac{\alpha^2
k^2}{2 m_{\rm{F}}}\frac{B_{j}}{A_{j}},\label{Om}
\\
 \label{AC}
{\mathcal{C}}_{0}^{2} = \frac{\alpha^2
B_{0}}{A_{0}}(A_{0}+B_{0})(A_{0}+B_{0} k^2),\qquad
{\mathcal{C}}_{j}^{2} = \frac{\alpha^2
B_{j}}{A_{j}}(A_{j}+B_{j})(A_{j}+B_{j} k^2),
\end{gather}
where $j=1,\ldots, N_{f}$. Next for convenience we introduce
\begin{gather*}
B_{0}=-\beta_{0} A_{0}, \qquad B_{j}=-\beta_{j} A_{j}, \qquad
j=1,\ldots , N_{f},
\end{gather*}
then
\begin{gather*} 
{\mathcal{C}}_{0}^{2} = \alpha^2 A_{0}^2\beta_{0}
(\beta_{0}-1)(1-\beta_{0} k^2),\qquad
{\mathcal{C}}_{j}^{2} = \alpha^2 A_{j}^2\beta_{j}
(\beta_{j}-1)(1-\beta_{j} k^2),\qquad j=1,\ldots, N_{f}.\nonumber
\end{gather*}
In order for our results (\ref{AbfV1}) to be consistent with the
parametrization (\ref{Abf1})--(\ref{Theta}) we must
ensure that both $q_{0}(x) $ and $\Theta_{0}(x) $ are real-valued,
and also $q_j(x) $ and $\Theta _j(x) $ are real-valued; this means
that $C_{0}^2\geq 0 $ and $q_{0}^2(x)\geq 0$ and also $C_j^2\geq 0
$ and $q_j^2(x)\geq 0 $ (see Table~\ref{tab:1}, $W_{\rm{B}}=(\alpha^2k^2-m_{\rm{B}}V_0)$,
$W_{\rm{F}}=(\alpha^2k^2-m_{\rm{F}}V_0)$). An elementary analysis
shows that with $l=0,\ldots, N_{f}$ one of the following conditions
must hold
\begin{gather*}
 \mbox{a)} \quad A_l\geq 0, \quad \beta _l\leq 0,  \qquad
 \mbox{b)} \quad A_l\leq 0, \quad 1\leq \beta
_l\leq {1\over k^2}.
\end{gather*}

\begin{table}[t]\centering
\caption{\label{tab:1}
$W=g_{\rm{BF}}m_{\rm{F}}W_{\rm{B}}/(m_{\rm{B}}W_{\rm{F}})$.}
\vspace{1mm}
\begin{tabular}{|l|l|l|l|l|l|l|l|} \hline
1  & $\beta_0\leq 0$ & $\beta_j\leq 0$ & $A_0\geq 0$ & $A_j\geq 0$
& $g_{\rm{BF}}\gtrless 0$ &
$g_{\rm{BB}}\lessgtr W$ &  $V_0\lessgtr \alpha^2 k^2/m_{\rm{F}}$ \\
\hline  2 & $\beta_0\leq 0$ & $1\leq\beta_j\leq 1/k^2$ & $A_0\geq
0$ & $A_j\leq 0$ &
$g_{\rm{BF}}\gtrless 0$ & $g_{\rm{BB}}\gtrless W$ & $V_0\lessgtr \alpha^2 k^2/m_{\rm{F}}$ \\
\hline  3  & $1\leq\beta_0\leq 1/k^2$ & $\beta_j\leq 0$ & $A_0\leq
0$ & $ A_j\geq 0$ &
 $g_{\rm{BF}}\gtrless 0$ & $g_{\rm{BB}}\gtrless W$ & $V_0 \gtrless \alpha^2 k^2/m_{\rm{F}}$ \\\hline
 4 & $1\leq\beta_0\leq 1/k^2$ & $1\leq\beta_j\leq 1/k^2$ & $A_0\leq 0$ & $A_j\leq 0$ &
$g_{\rm{BF}}\gtrless 0$  & $g_{\rm{BB}}\lessgtr W$ & $V_0 \gtrless
\alpha^2 k^2/m_{\rm{F}}$\\ \hline
\end{tabular}
\end{table}

Although our main interest is to analyze periodic solutions, note
that the solutions $\Psi^{b}$, $\Psi^{f}_{j}$ in (\ref{bf1}),
(\ref{bf2}) {\em are not} always periodic in $x$. Indeed, let us
f\/irst calculate explicitly $\Theta_{0}(x)$ and $\Theta_{j} (x)$ by
using the well known formula, see e.g.~\cite{as65}:
\begin{gather*}
 \int_{0}^{x}\frac{d u} {\wp(\alpha u)-\wp(\alpha v)}
 =\frac{1}{\wp'(\alpha v)}\left[2x\zeta(\alpha v)
+\frac{1}{\alpha } \ln \frac{\sigma(\alpha u-\alpha v)}
{\sigma(\alpha  u+\alpha v)} \right],
\end{gather*}
where $\wp$, $\zeta$, $\sigma$ are standard Weierstrass functions.

In the case a) we replace $v $ by $iv_{0} $ and $v $ by $iv_j $,
set $\sn^2(i\alpha v_{0};k)=\beta_{0}<0$, $\sn^2(i\alpha
v_j;k)=\beta_j<0$ and
\[
e_1=\frac13(2-k^2),\qquad  e_2=\frac13(2k^2-1),\qquad
e_3=-\frac13(1+k^2),
\]
and rewrite the l.h.s in terms of Jacobi elliptic functions:
\begin{gather*}
 \int_{0}^{x}\frac{du \;\sn^2(i\alpha v;k) \sn^2(\alpha u;k) }
{\sn^2(i\alpha v;k) - \sn^2(\alpha u;k)}= -\beta _{0}x -\beta
_{0}^2 \int_{0}^{x} \frac{d u\;}{\sn^2(\alpha u,k)-\beta _{0}},
\end{gather*}
and for $j=1,\ldots, N_{f}$ we have
\begin{gather*}
 \int_{0}^{x}\frac{du \;\sn^2(i\alpha v;k) \sn^2(\alpha u;k) }
{\sn^2(i\alpha v;k) - \sn^2(\alpha u;k)} = -\beta _jx -\beta _j^2
\int_{0}^{x} \frac{d u\;}{\sn^2(\alpha u,k)-\beta _j}.
\end{gather*}
Skipping the details we f\/ind the explicit form of
\begin{gather*}
\Theta _{0}(x) = C_{0} \int_{0}^{x} \frac{d u}
{A_{0}({\sn}^2(\alpha u;k)-\beta _{0}}= - \tau_{0} x + \frac{i}{2}
\ln \frac{\sigma(\alpha x+
i\alpha v_{0})} {\sigma(\alpha x-i\alpha v_{0})}, \\
\tau_{0} =  i\alpha \zeta (i\alpha v_{0})+ \frac{\alpha
}{\beta_{0}} \sqrt{-\beta_{0}(1-\beta_{0})(1-k^2\beta_{0})}.
\nonumber
\end{gather*}
and for  $\Theta _j(x)$, $j=1,\ldots, N_{f} $ we have
\begin{gather}\label{eq:Thet_j}
\Theta _j(x) = C_j \int_{0}^{x} \frac{d u} {A_j({\sn}^2(\alpha
u;k)-\beta _j)} = - \tau_j x +  \frac{i}{2} \ln
\frac{\sigma(\alpha x+
i\alpha v_j)} {\sigma(\alpha x-i\alpha v_j)}, \\
\tau_j =  i\alpha \zeta (i\alpha v_j)+ \frac{\alpha }{\beta_j}
\sqrt{-\beta_j(1-\beta_j)(1-k^2\beta_j)}. \nonumber
\end{gather}

These formulae provide an explicit expression for the solutions
$\Psi^{b}$, $\Psi^{f}_{j}$  with nontrivial phases; note that for
real values of $v_{0} $ $\Theta _{0}(x) $, $v_j $ $\Theta _j(x) $
are also real.  Now we can f\/ind the conditions under which
$Q_j(x,t) $ are periodic.  Indeed, from (\ref{eq:Thet_j}) we can
calculate the quantities $T_{0} $, $T_j $ satisfying:
\begin{gather*}
\Theta _{0}(x+T_{0})-\Theta _{0}(x) = 2\pi p_{0},\qquad
 \Theta_j(x+T_j)-\Theta _j(x) = 2\pi p_j,\qquad j=1,\ldots, N_{f}.
\end{gather*}
Then $\Psi^{b}$, $\Psi^{f}_{j}$  will be periodic in $x $ with
periods  $T_{0}=2m_{0}\omega /\alpha $, $T_j=2m_j\omega /\alpha $
if there exist pairs of integers $m_{0} $, $p_{0}$, and $m_j $,
$p_j$, such that:
\begin{gather*}
{m_{0} \over p_{0}} = -\pi \left[ \alpha v_{0}\zeta (\omega
)+\omega \tau_{0} /\alpha \right]^{-1},\qquad {m_j \over p_j} =
-\pi \left[ \alpha v_j\zeta (\omega )+\omega \tau_j /\alpha
\right]^{-1}, \qquad j=1, \ldots, N_{f}.\nonumber
\end{gather*}
where $\omega $ (and  $\omega' $) are the half-periods of the
Weierstrass functions.

\section{Type B nontrivial phase solutions}\label{sec:3}

For the f\/irst time solutions of this type were derived in
\cite{bcdn01,ckr01,bcdkp01} for the case of nonlinear
Schr\"{o}dinger equation  and in \cite{Deco} for the $n$-component
CNLSE. For Bose--Fermi mixtures solutions of this type are possible
\begin{itemize}\itemsep=0pt
\item  when we have two lattices $V_{\rm{B}}$ and $V_{\rm{F}}$,

\item when $m_{\rm{B}}=m_{\rm{F}}$.

\end{itemize}
We seek the solutions in one of the following forms:
\begin{alignat}{3}
& q_{0}^{2}= A_{0}\sn(\alpha x,k)+B_{0}, \qquad  && q_{j}^{2}=
A_{j}\sn(\alpha x,k)+B_{j},\label{AbfV1B1} & \\
& q_{0}^{2}= A_{0}\cn(\alpha x,k)+B_{0}, \qquad && q_{1}^{2}=
A_{j}\cn(\alpha x,k)+B_{j},\label{AbfV1B2}&\\
& q_{0}^{2}= A_{0}\dn(\alpha x,k)+B_{0}, \qquad  && q_{1}^{2}=
A_{j}\dn(\alpha x,k)+B_{j},\qquad j=1,\ldots, N_{f}.\label{AbfV1B3}&
\end{alignat}
In the f\/irst case (\ref{AbfV1B1}) we have
\begin{gather*}
V_{\rm{B}}=\frac{3\alpha^2 k^2}{8 m_{\rm{B}}},\qquad
V_{\rm{F}}=\frac{3\alpha^2 k^2}{8 m_{\rm{F}}}
\\
A_{0}=-\frac{\alpha^2 k^2}{4 m_{\rm{F}} g_{\rm{BF}}
}\frac{B_{j}}{A_{j}},\qquad
\frac{B_1}{A_1}=\dots=\frac{B_{N_f}}{A_{N_F}}, \qquad \sum_j
A_{j}=-\frac{\alpha^2 k^2}{4 m_{\rm{B}}g_{\rm{BF}}}
\frac{B_{0}}{A_{0}}-\frac{A_{0} g_{\rm{BB}} }{g_{\rm{BF}}},
\\
\omega_{0}=\frac{\alpha^2 (k^2+1)}{8 m_{\rm{B}}}
+g_{\rm{BB}}B_{0}+g_{\rm{BF}} B_{1} -\frac{\alpha^2k^2}{8
m_{\rm{B}}} \frac{B^2_{0}}{A^2_{0}},\nonumber\\
\omega_{j}=\frac{\alpha^2 (k^2+1)}{8 m_{\rm{F}}}+ g_{\rm{BF}}
B_{0} -\frac{\alpha^2 k^2}{8 m_{\rm{F}}} \frac{B^2_{j}}{A^2_{j}},
\\ 
{\mathcal{C}}_{0}^{2} = \frac{\alpha^2 }{4
A^2_{0}}(B_{0}^2-A_{0}^2)(A_{0}^2-B_{0}^2 k^2),\qquad
{\mathcal{C}}_{j}^{2} = \frac{\alpha^2}{4
A_{j}^2}(B_{j}^2-A_{j}^2)(A_{j}^2-B_{j}^2 k^2).
\end{gather*}
We remark that due to relations
$\frac{B_1}{A_1}=\dots=\frac{B_{N_f}}{A_{N_F}}$ we have that all
$q_j$ of the fermion f\/ields are proportional to $q_{1}$.

\section{Examples of elliptic solutions}\label{sec:4}

Using the general solution  equations (\ref{A})--(\ref{AC}) we have the
following special cases: (these solutions are possible only when
we have some restrictions on $g_{\rm{BB}}$, $g_{\rm{BF}}$, and
$V_{0}$ see the Table~\ref{tab:1})

\begin{example} Suppose that $B_0=B_j=0$. Therefore we have
\begin{gather}
q_0(x)=\sqrt{A_0}\sn(\alpha x,k),\qquad q_j=\sqrt{A_j}\sn(\alpha
x,k), \label{Ell1}
\\
\label{A_0}
A_0=\frac{\alpha^2 k^2-m_{\rm{F}}V_0}{m_{\rm{F}}g_{\rm{BF}}},
\qquad 
\sum_jA_j=\frac{\alpha^2
k^2}{g_{\rm{BF}}}\left(\frac{1}{m_{\rm{B}}}-\frac{g_{\rm{BB}}}{m_Fg_{\rm{FB}}}\right)
-\frac{V_0}{g_{\rm{BF}}}\left(1-\frac{g_{\rm{BB}}}{g_{\rm{BF}}}\right).
\end{gather}
 For the frequencies $\omega_0$ and $\omega_j$ we have
\begin{gather*}
\omega_0=\frac{\alpha^2(1+k^2)}{2m_{\rm{B}}},\qquad
\omega_j=\frac{\alpha^2(1+k^2)}{2m_{\rm{F}}}.
\end{gather*}
as well as $\mathcal{C}_0=\mathcal{C}_j=0$.
\end{example}

\begin{example}
Let $B_0=-A_0$ and $B_j=-A_j$ hold true. Then we have
\begin{gather}
q_0(x)=\sqrt{-A_0}\cn(\alpha x,k),\qquad
q_j(x)=\sqrt{-A_j}\cn(\alpha x,k).
\end{gather}
The coef\/f\/icients $A_0$ and $A_j$ have the same form as (\ref{A_0}).
 The frequencies $\omega_0$ and $\omega_j$ now
look as follows
\begin{gather*}
\omega_0=\frac{\alpha^2(1-2k^2)}{2m_{\rm{B}}}+V_0,\qquad
\omega_j=\frac{\alpha^2(1-2k^2)}{2m_{\rm{F}}}+V_0.
\end{gather*}
The constants $\mathcal{C}_0$ and $\mathcal{C}_j$ are equal to
zero again.
\end{example}

\begin{example}
$B_0=-A_0/k^2$ and $B_j=-A_j/k^2$. In this case we obtain
\begin{gather}
q_0(x)=\frac{\sqrt{-A_0}}{k}\dn(\alpha x,k),\qquad
q_j(x)=\frac{\sqrt{-A_j}}{k}\dn(\alpha x,k),\nonumber
\\
\omega_0=\frac{\alpha^2(k^2-2)}{2m_{\rm{B}}}+\frac{V_0}{k^2},\qquad
\omega_j=\frac{\alpha^2(k^2-2)}{2m_{\rm{F}}}+\frac{V_0}{k^2}.
\end{gather}
As before $\mathcal{C}_0=\mathcal{C}_j=0$.
\end{example}

\begin{example}
$B_0=0$ and $B_j=-A_j$. The result reads
\begin{gather}
q_0(x)=\sqrt{A}_0\sn(\alpha x,k),\qquad  q_j(x)=\sqrt{-A_j}\cn(\alpha x,k),\nonumber\\
\omega_0=\frac{\alpha^2(1-k^2)}{2m_{\rm{B}}}+V_0+A_0g_{\rm{BB}},\qquad
\omega_j=\frac{\alpha^2}{2m_{\rm{F}}}.
\end{gather}
By analogy with the previous examples the constants $A_0$, $A_j$,
$\mathcal{C}_0$ and $\mathcal{C}_j$ are given by formu\-lae~(\ref{A_0}) 
and $\mathcal{C}_0$, $\mathcal{C}_j$
are all zero.
\end{example}

\begin{example}
$B_0=0$ and $B_j=-A_j/k^2$. Thus one gets
\begin{gather}
 q_0(x)=\sqrt{A_0}\sn(\alpha x,k),\qquad  q_j(x)=\frac{\sqrt{-A_j}}{k}\dn(\alpha x,k),\nonumber\\
\omega_0=\frac{\alpha^2(k^2-1)}{2m_{\rm{B}}}+\frac{V_0}{k^2}+\frac{A_0g_{\rm{BB}}}{k^2},\qquad
\omega_j=\frac{\alpha^2k^2}{2m_{\rm{F}}}.\label{Ell11}
\end{gather}
\end{example}

\begin{example}
Let $B_0=-A_0$ and $B_j=0$. Hence we have
\begin{gather*}
 q_0(x)=\sqrt{-A_0}\cn(\alpha x,k),\qquad q_j(x)=\sqrt{A_j}\sn(\alpha x,k),\\
 \omega_0=\frac{\alpha^2}{2m_{\rm{B}}}-g_{\rm{BB}}A_0,\qquad
\omega_j=\frac{\alpha^2(1-k^2)}{2m_{\rm{F}}}+V_0.
\end{gather*}
\end{example}

\begin{example}
Let $B_0=-A_0$ and $B_j=-A_j/k^2$. We obtain
\begin{gather*}
 q_0(x)=\sqrt{-A_0}\cn(\alpha x,k),\qquad  q_j(x)=\frac{\sqrt{-A_j}}{k}\dn(\alpha x,k),\\
\omega_0=\frac{V_0}{k^2}-\frac{\alpha^2}{2m_{\rm{B}}}+\frac{1-k^2}{k^2}A_0g_{\rm{BB}},\qquad
\omega_j=V_0-\frac{\alpha^2k^2}{2m_{\rm{F}}}.
\end{gather*}
\end{example}

\begin{example}
Suppose $B_0=-A_0/k^2$ and $B_j=0$. Then
\begin{gather*}
 q_0(x)=\frac{\sqrt{-A_0}}{k}\dn(\alpha x,k), \qquad q_j(x)=\sqrt{A_j}\sn(\alpha x,k),\\
\omega_0=\frac{\alpha^2k^2}{2m_{\rm{B}}}-\frac{g_{\rm{BB}}A_0}{k^2},\qquad
\omega_j=\frac{\alpha^2(k^2-1)}{2m_{\rm{F}}}+\frac{V_0}{k^2}.
\end{gather*}
\end{example}

\begin{example}
Let $B_0=-A_0/k^2$ and $B_j=-A_j$. Thus
\begin{gather*}
 q_0(x)=\frac{\sqrt{-A_0}}{k}\dn(\alpha x,k), \qquad q_j(x)=\sqrt{-A_j}\cn(\alpha x,k),\\
\omega_0=V_0-\frac{\alpha^2k^2}{2m_{\rm{B}}}+\frac{k^2-1}{k^2}g_{\rm{BB}}A_0,\qquad
\omega_j=\frac{V_0}{k^2}-\frac{\alpha^2}{2m_{\rm{F}}}.
\end{gather*}
\end{example}
All these cases when $V_{0}=0$ and $j=2$ are derived for the f\/irst
time in~\cite{bpv07}.
\begin{table}[t]\centering
\caption{\label{tab:2} Trivial phase solutions in the generic
case. We use the quantity
$W=g_{\rm{BF}}m_{\rm{F}}W_{\rm{B}}/(m_{\rm{B}}W_{\rm{F}})$.}

\vspace{1mm}

\begin{tabular}{|l|l|l|l|l|} \hline
 1  & $q_0=\sqrt{A_0}\sn (\alpha x,k)$ & $g_{\rm{BF}}\gtrless 0$ & $g_{\rm{BB}}\lessgtr W$
&  $V_0\lessgtr \alpha^2 k^2/m_{\rm{F}}$\\
& $ q_j=\sqrt{A_j}\sn (\alpha x,k)$ &  &  &  \\
\hline  2  & $ q_0 =\sqrt{-A_0}\cn (\alpha
x,k) $ & $g_{\rm{BF}}\gtrless 0$ & $g_{\rm{BB}}\lessgtr W$ &  $V_0 \gtrless \alpha^2 k^2/m_{\rm{F}}$\\
 & $ q_j=\sqrt{-A_j} \cn (\alpha x,k) $ & & & \\
\hline  3  & $ q_0 =\sqrt{-A_0}\dn (\alpha x,k)/k$ &
$g_{\rm{BF}}\gtrless 0$ & $g_{\rm{BB}}\lessgtr W$
& $V_0 \gtrless \alpha^2 k^2/m_{\rm{F}}$ \\
& $q_j=\sqrt{-A_j}\dn (\alpha x,k)/k$ &  &  &  \\
\hline  4 & $q_0 =\sqrt{A_0} \sn (\alpha x,k)$ &
$g_{\rm{BF}}\gtrless 0$ & $g_{\rm{BB}}\gtrless W$
& $V_0\lessgtr \alpha^2 k^2/m_{\rm{F}}$ \\
& $ q_j=\sqrt{-A_j} \cn (\alpha x,k)$ &  &  &  \\
\hline  5 & $q_0 =\sqrt{A_0} \sn (\alpha x,k)$ & $g_{\rm{BF}}\gtrless 0$ &
$g_{\rm{BB}}\gtrless W$ & $V_0\lessgtr \alpha^2 k^2/m_{\rm{F}}$\\
& $ q_j=\sqrt{-A_j}\dn (\alpha x,k)/k$ &  &  & \\
\hline  6 & $q_0 =\sqrt{-A_0}\cn (\alpha x,k)$ &
$g_{\rm{BF}}\gtrless 0$ & $g_{\rm{BB}}\gtrless W$
& $V_0 \gtrless \alpha^2 k^2/m_{\rm{F}}$ \\
& $ q_j=\sqrt{A_j} \sn (\alpha x,k)$ &  &  &  \\ \hline
 7 & $q_0 =\sqrt{-A_0} \cn (\alpha x,k)$ & $g_{\rm{BF}}\gtrless 0$  & $g_{\rm{BB}}\lessgtr W$
& $V_0 \gtrless \alpha^2 k^2/m_{\rm{F}}$ \\
& $ q_j=\sqrt{-A_j} \dn (\alpha x,k)/k$ & & & \\ \hline
 8 & $q_0 =\sqrt{-A_0} \dn (\alpha x,k)/k$ & $g_{\rm{BF}}\gtrless 0$ & $g_{\rm{BB}}\gtrless W$
& $V_0 \gtrless \alpha^2 k^2/m_{\rm{F}}$ \\
& $q_j=\sqrt{A_j} \sn (\alpha x,k)$ & &  &  \\ \hline
 9 & $q_0 =\sqrt{-A_0} \dn (\alpha x,k)/k$ & $g_{\rm{BF}}\gtrless 0$ & $g_{\rm{BB}}\lessgtr W$ &
$V_0 \gtrless \alpha^2 k^2/m_{\rm{F}}$\\
& $q_j=\sqrt{-A_j} \cn (\alpha x,k)$ & & & \\\hline
\end{tabular}
\end{table}

\subsection{Mixed trivial phase solution}

\begin{example}
When $B_{0}=0$, $B_{1}=0$, $B_{2}=-A_{2}$, $B_{j}=- A_{j}/k^2$,
$j=3,\ldots, N_{f}$ the solutions obtain the form
\begin{gather*}
 q_{0}= \sqrt{A_{0}}\sn(\alpha x,k), \qquad  q_{1}=\sqrt{A_{1}}
\sn(\alpha x,k),\nonumber\\
 q_{2}=\sqrt{-A_{2}} \cn(\alpha
x,k),\qquad q_{j}=\sqrt{-A_{j}}\dn(\alpha x,k)/k,\qquad j=3,\ldots,
N_{f}. \nonumber
\end{gather*}
Using equations (\ref{A})--(\ref{AC}) we have
\begin{gather*}
A_{0}=\frac{\alpha^2 k^2-V_{0}m_{F}}{m_{F} g_{\rm{BF}}},\qquad
\sum_{j=1}^{N_{f}} A_{j}=\alpha^2 k^2\left(\frac{1}{m_{B}
g_{\rm{BF}}}-\frac{g_{\rm{BB}}}{m_{F} g^2_{\rm{BF}}}\right)-
V_{0}\left(\frac{1}{g_{\rm{BF}}}-\frac{g_{\rm{BB}}}{g^2_{\rm{BF}}}\right),\\
\omega_{0}=\frac{\alpha^2 (k^2-1)}{2 m_{B}}
+\frac{g_{\rm{BF}}}{k^2}\left(A_{1}+(1-k^2) A_{2}
\right)+\frac{g_{\rm{BB}}A_{0}}{k^2}+\frac{V_{0}}{k^2},\\
\omega_{1}=\frac{\alpha^2 (1+k^2)}{2 m_{F}}, \qquad
\omega_{2}=\frac{1}{2 m_{F}}\alpha^2, \qquad
\omega_{j}=\frac{\alpha^2 k^2}{2 m_{F}}, \qquad j=3,\ldots, N_{F}.
\end{gather*}
\end{example}

\begin{example}
Let $B_0=B_1=0$ and $B_j=-A_j$ where $j=2,\ldots,N_f$. Therefore
the solutions read
\begin{gather*}
q_0(x)=\sqrt{A_0}\sn(\alpha x,k),\qquad
q_1(x)=\sqrt{A_1}\sn(\alpha x,k),\qquad
q_j(x)=\sqrt{-A_j}\cn(\alpha x,k).
\end{gather*}
Then we obtain for frequencies the following results
\begin{gather*}
\omega_0=\frac{\alpha^2(1-k^2)}{2m_{\rm{B}}}+V_0+g_{\rm{BB}}A_0+g_{\rm{BF}}A_1,
\qquad \omega_1=\frac{\alpha^2(1+k^2)}{2m_{\rm{F}}},\qquad
\omega_j=\frac{\alpha^2}{2m_{\rm{F}}}.
\end{gather*}
\end{example}

\begin{example}
Suppose $B_0=-A_0$, $B_1=0$, $B_2=-A_2$ and $B_j=-A_j/k^2$ where
$j=3,\ldots,N_f$. The solutions have the form
\begin{gather*}
q_0(x)=\sqrt{-A_0}\cn(\alpha x,k),\qquad q_1(x)=\sqrt{A_1}\sn(\alpha x,k),\\
q_2(x)=\sqrt{-A_2}\cn(\alpha x,k),\qquad
q_j(x)=\sqrt{-A_j}\dn(\alpha x,k)/k.
\end{gather*}
The frequencies are
\begin{gather*}
\omega_0=\frac{V_0}{k^2}-\frac{\alpha^2}{2m_{\rm{B}}}+\frac{1-k^2}{k^2}
(g_{\rm{BB}}A_0+g_{\rm{BF}}A_2)+\frac{g_{\rm{BF}}}{k^2}A_1,\qquad
\omega_1=V_0+\frac{\alpha^2(1-k^2)}{2m_{\rm{F}}},\\
\omega_2=V_0+\frac{\alpha^2(1-2k^2)}{2m_{\rm{F}}},\qquad
\omega_j=V_0-\frac{\alpha^2k^2}{2m_{\rm{F}}}.
\end{gather*}
\end{example}

\begin{example}
Let $B_0=-A_0$, $B_1=-A_1$ and $B_j=-A_j/k^2$ for
$j=2,\ldots,N_f$. Then
\begin{gather*}
q_0(x)=\sqrt{-A_0}\cn(\alpha x,k),\qquad
q_1(x)=\sqrt{-A_1}\cn(\alpha x,k),\qquad
q_j(x)=\sqrt{-A_j}\dn(\alpha x,k)/k,
\\
\omega_0=\frac{V_0}{k^2}-\frac{\alpha^2}{2m_{\rm{B}}}+\frac{1-k^2}{k^2}
\left(g_{\rm{BB}}A_0+g_{\rm{BF}}A_1\right),\qquad \omega_1=V_0+
\frac{\alpha^2(1-2k^2)}{2m_{\rm{F}}},\nonumber\\
\omega_j=V_0-\frac{\alpha^2k^2}{2m_{\rm{F}}}.
\end{gather*}
\end{example}
\begin{example}
Let $B_0=-A_0/k^2$, $B_1=-A_1$ and $B_j=-A_j/k^2$ for
$j=2,\ldots,N_f$. Hence
\begin{gather*}
q_0(x)=\sqrt{-A_0}\dn(\alpha x,k)/k,\quad
q_1(x)=\sqrt{-A_1}\cn(\alpha x,k),\qquad
q_j(x)=\sqrt{-A_j}\dn(\alpha x,k)/k,
\\
\omega_0=\frac{\alpha^2(k^2-2)}{2m_{\rm{B}}}+\frac{V_0}{k^2}
+\frac{1-k^2}{k^2}(g_{\rm{BB}}A_0+g_{\rm{BF}}A_1),\\ \omega_1=
\frac{V_0}{k^2}-\frac{\alpha^2}{2m_{\rm{F}}},\qquad
\omega_j=\frac{V_0}{k^2}+\frac{\alpha^2(k^2-2)}{2m_{\rm{F}}}.
\end{gather*}
\end{example}
Certainly these examples do not exhaust all possible combinations
of solutions and it is easy to extend this list.

\section{Vector soliton solutions}\label{sec:5}
\subsection{Vector bright-bright soliton solutions}

When $k\rightarrow1$, $\sn(\alpha x,1)=\tanh(\alpha x)$ and
$B_{0}=-A_{0}$, $B_{j}=-A_{j}$ we obtain that the solutions read
\begin{gather*}
q_{0}= \sqrt{-A_{0}}\frac{1}{\cosh(\alpha x)}, \qquad  q_{j}=
\sqrt{-A_{j}}\frac{1}{\cosh(\alpha x)},
\end{gather*}
where $A_0\leq 0$ as well as $A_j\leq 0$. Using equations (\ref{A})--(\ref{AC}) we have
\begin{gather*}
A_{0}=\frac{\alpha^2-V_{0}m_{\rm{F}}}{m_{\rm{F}} g_{\rm{BF}}},
\qquad V=V_{0}\tanh^2(\alpha x) ,
\\
\sum_{j=1}^{N_{f}} A_{j}=\frac{\alpha^2}{g_{\rm{BF}}}
\left(\frac{1}{m_{\rm{B}}}-\frac{g_{\rm{BB}}}{m_{\rm{F}}g_{\rm{BF}}}\right)-
\frac{V_{0}}{g_{\rm{BF}}}\left(1-\frac{g_{\rm{BB}}}{g_{\rm{BF}}}\right),
\\ 
\omega_{0}=V_0-\frac{1}{2 m_{B}}\alpha^2, \qquad
\omega_{j}=V_0-\frac{1}{2 m_{F}}\alpha^2.
\end{gather*}
As a consequence of the restrictions on $A_0$ and $A_j$ one can
get the following unequalities
\begin{gather*}
g_{\rm{BF}}> 0,\qquad V_0\geq\frac{\alpha^2}{m_{\rm{F}}}, \qquad
g_{\rm{BB}}\leq\frac{(\alpha^2-m_{\rm{B}}V_0)m_{\rm{F}}}
{(\alpha^2-m_{\rm{F}}V_0)m_{\rm{B}}}g_{\rm{BF}},\\
g_{\rm{BF}}< 0,\qquad V_0\leq \frac{\alpha^2}{m_{\rm{F}}}, \qquad
g_{\rm{BB}}\geq \frac{(\alpha^2-m_{\rm{B}}V_0)m_{\rm{F}}}
{(\alpha^2-m_{\rm{F}}V_0)m_{\rm{B}}}g_{\rm{BF}}.
\end{gather*}
Vector bright soliton solution when $V_{0}=0$ is derived for the
f\/irst time in \cite{bpv07}.

\subsection{Vector dark-dark soliton solutions}

When $k\rightarrow1$ and $B_0=B_j=0$ are satisf\/ied the solutions
read
\begin{gather*}
q_0(x)=\sqrt{A_0}\tanh(\alpha x),\qquad
q_j(x)=\sqrt{A_j}\tanh(\alpha x). 
\end{gather*}
The natural restrictions $A_0\geq 0$ and $A_j\geq 0$ lead to
\begin{gather}
g_{\rm{BF}}> 0,\qquad g_{\rm{BB}}\leq
\frac{(\alpha^2-m_{\rm{B}}V_0)m_{\rm{F}}}
{(\alpha^2-m_{\rm{F}}V_0)m_{\rm{B}}}g_{\rm{BF}},\qquad
V_0\leq \alpha^2 /m_{\rm{F}},\nonumber\\
g_{\rm{BF}}< 0,\qquad g_{\rm{BB}}\geq
\frac{(\alpha^2-m_{\rm{B}}V_0)m_{\rm{F}}}
{(\alpha^2-m_{\rm{F}}V_0)m_{\rm{B}}}g_{\rm{BF}},\qquad V_0\geq
\alpha^2 /m_{\rm{F}},\nonumber\\
\label{vsol1_A}
A_0=\frac{\alpha^2 -m_{\rm{F}}V_0}{m_{\rm{F}}g_{\rm{BF}}},\qquad
\sum_jA_j=\frac{\alpha^2}{g_{\rm{BF}}}\left(\frac{1}{m_{\rm{B}}}
-\frac{g_{\rm{BB}}}{m_Fg_{\rm{FB}}}\right)
-\frac{V_0}{g_{\rm{BF}}}\left(1-\frac{g_{\rm{BB}}}{g_{\rm{BF}}}\right).
\end{gather}
For the frequencies $\omega_0$ and $\omega_j$ and the constants
$\mathcal{C}_0$ and $\mathcal{C}_j$ we have
\begin{gather}
\omega_0=\frac{\alpha^2}{m_{\rm{B}}},\qquad
\omega_j=\frac{\alpha^2}{m_{\rm{F}}},\qquad
\mathcal{C}_0=\mathcal{C}_j=0.\label{vsol1_C}
\end{gather}

\subsection{Vector bright-dark soliton solutions}

When $k\rightarrow 1$, $B_0=-A_0$ and $B_j=0$, we have
\begin{gather*}
 q_0(x)=\frac{\sqrt{-A_0}}{\cosh(\alpha x)},\qquad q_j(x)=\sqrt{A_j}\tanh(\alpha x),\\
 \omega_0=\frac{\alpha^2}{2m_{\rm{B}}}-g_{\rm{BB}}A_0,\qquad
\omega_j=V_0,\qquad \mathcal{C}_0=\mathcal{C}_j=0.
\end{gather*}
The parameters $A_0$ and $A_j$ are given by (\ref{vsol1_A}). In
this case we have the following restrictions
\begin{gather*}
g_{\rm{BF}}> 0,\qquad g_{\rm{BB}}\geq
\frac{(\alpha^2-m_{\rm{B}}V_0)m_{\rm{F}}}
{(\alpha^2-m_{\rm{F}}V_0)m_{\rm{B}}}g_{\rm{BF}},\qquad V_0
\geq \alpha^2 /m_{\rm{F}},\\
g_{\rm{BF}}< 0,\qquad g_{\rm{BB}}\leq
\frac{(\alpha^2-m_{\rm{B}}V_0)m_{\rm{F}}}
{(\alpha^2-m_{\rm{F}}V_0)m_{\rm{B}}}g_{\rm{BF}},\qquad V_0 \leq
\alpha^2 /m_{\rm{F}}.
\end{gather*}

\subsection{Vector dark-bright soliton solutions}

When $k\rightarrow1$ and provided that $B_0=0$ and $B_j=-A_j$ the
result is
\begin{gather*}
 q_0(x)=\sqrt{A}_0\tanh(\alpha x),\qquad
q_j(x)=\frac{\sqrt{-A_j}}{\cosh(\alpha x)},\qquad
\omega_0=V_0+A_0g_{\rm{BB}},\qquad
\omega_j=\frac{\alpha^2}{2m_{\rm{F}}}.
\end{gather*}
By analogy with the previous examples the constants $A_0$, $A_j$,
$\mathcal{C}_0$ and $\mathcal{C}_j$ are given by formu\-lae~(\ref{vsol1_A}) and (\ref{vsol1_C}) respectively. The restrictions
now are
\begin{gather*}
g_{\rm{BF}}> 0,\qquad g_{\rm{BB}}\geq
\frac{(\alpha^2-m_{\rm{B}}V_0)m_{\rm{F}}}
{(\alpha^2-m_{\rm{F}}V_0)m_{\rm{B}}}g_{\rm{BF}},\qquad
V_0\leq \alpha^2 /m_{\rm{F}},\\
g_{\rm{BF}}< 0,\qquad g_{\rm{BB}}\leq
\frac{(\alpha^2-m_{\rm{B}}V_0)m_{\rm{F}}}
{(\alpha^2-m_{\rm{F}}V_0)m_{\rm{B}}}g_{\rm{BF}},\qquad V_0\geq
\alpha^2 /m_{\rm{F}}.
\end{gather*}

\subsection{Vector dark-dark-bright soliton solutions}
Let $B_0=B_1=0$ and $B_j=-A_j$ where $j=2,\ldots,N_f$. Therefore
the solutions read
\begin{gather*}
q_0(x)=\sqrt{A_0}\tanh(\alpha x),\qquad
q_1(x)=\sqrt{A_1}\tanh(\alpha x),\qquad
q_j(x)=\sqrt{-A_j}\sech(\alpha x).
\end{gather*}
Then we obtain for frequencies the following results
\begin{gather*}
\omega_0=V_0+g_{\rm{BB}}A_0+g_{\rm{BF}}A_1, \qquad
\omega_1=\frac{\alpha^2}{m_{\rm{F}}},\qquad
\omega_j=\frac{\alpha^2}{2m_{\rm{F}}}.
\end{gather*}
These examples are by no means exhaustive.

\subsection{Nontrivial phase, trigonometric limit}

In this section we consider a trap potential of the form
$V_{\rm{trap}} = V_{0} \cos(2\alpha x)$, as a model for an optical
lattice. Our potential $V$ is similar and dif\/fers only with
additive constant.  When $k\rightarrow 0$, $\sn(\alpha
x,0)=\sin(\alpha x)$
\begin{gather}
q_{0}^{2}= A_{0}\sin^{2}(\alpha x)+B_{0},\qquad q_{j}^{2}=
A_{j}\sin^{2}(\alpha x)+B_{j},\label{SSol1}\\
V=V_{0}\sin^2(\alpha x)=\frac{1}{2}(V_{0}-V_{0}\cos(2\alpha x)),
\label{SSol2}
\end{gather}
Using equations (\ref{A})--(\ref{AC}) again we obtain the following
result when (see Table~\ref{tab:2a})
\begin{gather*}
A_{0}=-\frac{V_{0}}{g_{\rm{BF}}}, \qquad \sum_{j=1}^{N_{f}}
A_{j}=-\frac{V_{0}}{g_{\rm{BF}}}\left(1-\frac{g_{\rm{BB}}}{g_{\rm{BF}}}\right),\\
\omega_{0}=\frac{1}{2 m_{\rm{B}}}\alpha^2+B_{0} g_{\rm{BB}}+
g_{\rm{BF}}\sum_{i=1}^{N_{f}} B_{i} ,\qquad \omega_{j}=\frac{1}{2
m_{\rm{F}}}\alpha^2+g_{\rm{BF}} B_{0},\\
{\mathcal{C}}_{0}^{2} = \alpha^2 B_{0}(A_{0}+B_{0}),\qquad
{\mathcal{C}}_{j}^{2} = \alpha^2 B_{j}(A_{j}+B_{j}),
\end{gather*}
where
\begin{gather*}
\Theta_{0}(x)=\arctan\left(\sqrt{\frac{A_{0}+ B_{0}}
{B_{0}}}\tan(\alpha x)\right), \quad \Theta
_{j}(x)=\arctan\left(\sqrt{\frac{A_{j}+ B_{j}}{B_{j}}}\tan(\alpha
x)\right).\nonumber
\end{gather*}

\begin{table}[t]\centering
\caption{\label{tab:2a}
$W=g_{\rm{BF}}m_{\rm{F}}W_{\rm{B}}/(m_{\rm{B}}W_{\rm{F}})$.}
\vspace{1mm}

\begin{tabular}{|l|l|l|l|l|l|l|l|} \hline
1  & $\beta_0\leq 0$ & $\beta_j\leq 0$ & $A_0\geq 0$ & $A_j\geq 0$
& $g_{\rm{BF}}\gtrless 0$ &
$g_{\rm{BB}}\lessgtr g_{\rm{BF}}$ &  $V_0\lessgtr 0$ \\
\hline  2 & $\beta_0\leq 0$ & $\beta_j\geq 1$ & $A_0\geq 0$ &
$A_j\leq 0$ &
$g_{\rm{BF}}\gtrless 0$ & $g_{\rm{BB}}\gtrless g_{\rm{BF}}$ & $V_0\lessgtr 0$ \\
\hline  3  & $\beta_0\geq 1$ & $\beta_j\leq 0$ & $A_0\leq 0$ & $
A_j\geq 0$ &
 $g_{\rm{BF}}\gtrless 0$ & $g_{\rm{BB}}\gtrless g_{\rm{BF}}$ & $V_0 \gtrless 0$ \\\hline
 4 & $\beta_0\geq 1$ & $\beta_j\geq 1$ & $A_0\leq 0$ & $A_j\leq 0$ &
$g_{\rm{BF}}\gtrless 0$  & $g_{\rm{BB}}\lessgtr g_{\rm{BF}}$ &
$V_0 \gtrless 0$\\ \hline
\end{tabular}
\end{table}

This solution is the most important from the physical point of
view \cite{s05}.

\section{Linear stability, preliminary results}\label{sec:7}
To analyze linear stability of our initial system of equations we
seek solutions in the form
\begin{gather*}
\psi_0(x,t)=(q_0(x)+\varepsilon\phi_0(x,t))\exp\left(-\frac{i\omega_0}{\hbar}t+i\Theta_0(x)+i\kappa_0\right),\\
\psi_j(x,t)= (q_1(x)+\varepsilon\phi_j
(x,t))\exp\left(-\frac{i\omega_j}{\hbar}t+i\Theta_1(x)+i\kappa_1\right).
\end{gather*}
and obtain the following linearized equations
\begin{gather*}
\hbar\left(\begin{array}{c}
{\bf \Phi}_0\\{\bf \Phi}_1\\
\vdots\\{\bf \Phi}_{N_f}
\end{array}\right)_{,t}=\left(\begin{array}{ccccc}
{\bf\Lambda}_0 & {\bf U_1}      & {\bf U_2}      & \hdots & {\bf U_{N_f}} \\
{\bf V_1}      & {\bf\Lambda}_1 & 0              & \hdots & 0             \\
{\bf V_2}      & 0              & {\bf\Lambda}_2 & \hdots & 0             \\
 \vdots        & \vdots         & \vdots         & \ddots & \vdots        \\
{\bf V_{N_f}}  & 0              & 0              & \hdots &
{\bf\Lambda}_{N_f}
\end{array}\right)\left(\begin{array}{c}
{\bf \Phi}_0\\{\bf \Phi}_1\\
\vdots\\{\bf \Phi}_{N_f}\\
\end{array}\right),\qquad\!
{\bf \Phi}_0=\left(\begin{array}{c} \phi^{\rm{R}}_0\\
\phi^{\rm{I}}_0,
\end{array}\right), \qquad\!
{\bf\Phi}_j=\left(\begin{array}{c} \phi^{\rm{R}}_j\\
\phi^{\rm{I}}_j
\end{array}\right),
\end{gather*}
where
\begin{gather*}
{\bf\Lambda}_{0}=\left(\begin{array}{cc}
S_0       & L_{0,-}\\
L_{0,+}   & S_0
\end{array}\right),\quad
{\bf U_j}=\left(\begin{array}{cc}
 0       & 0 \\
 U_{0,j} &  0
\end{array}\right),\quad
{\bf\Lambda}_{j}=\left(\begin{array}{cc}
S_j       & L_{j,-}\\
L_{j,+}   & S_j
\end{array}\right),\quad
{\bf V_j}=\left(\begin{array}{cc}
0       & 0 \\
 U_{1,j} &  0
\end{array}\right), \\
S_0=-\frac{\mathcal{C}_0}{m_{\rm{B}}q_0}\partial_x\left(\frac{1}{q_0}\right),\qquad
L_{0,-}=-\frac{1}{2m_{\rm{B}}}\left(\partial^2_{xx}-\frac{\mathcal{C}^2_0}{q^4_0}\right)
+V+g_{\rm{BB}}q^2_0+g_{\rm{BF}}q^2_1-\omega_0, \\
U_{0,j}=-2g_{\rm{BF}}q^2_0,\qquad
L_{0,+}=\frac{1}{2m_{\rm{B}}}\left(\partial^2_{xx}
-\frac{\mathcal{C}^2_0}{q^4_0}\right)- V -3g_{\rm{BB}}q^2_0-g_{\rm{BF}}q^2_1+\omega_0, \\
S_j=-\frac{\mathcal{C}_j}{m_{\rm{F}}q_j}\partial_x\left(\frac{1}{q_j}\right),\qquad
L_{j,-}=
-\frac{1}{2m_{\rm{F}}}\left(\partial^2_{xx}-\frac{\mathcal{C}^2_j}{q^4_0}\right)
+V+g_{\rm{BF}}q^2_0-\omega_j, \\
U_{1,j}=-2g_{\rm{BF}}q_0q_j,\qquad L_{j,+}=
\frac{1}{2m_{\rm{F}}}\left(\partial^2_{xx}-\frac{\mathcal{C}^2_j}{q^4_0}\right)
- V - g_{\rm{BF}}q^2_0+\omega_j, \qquad j=1,\ldots, N_{f}.
\end{gather*}
The analysis of the latter matrix system is a dif\/f\/icult problem and
only numerical simulations are possible.  Recently
a great progress was achieved for analysis of linear stability of periodic
solutions of type (\ref{Abf1}), (\ref{Abf2}) (see e.g.
\cite{bcdn01,ckr01,bcdkp01,Deco,kegks04} and references therein).
Nevertheless the stability analysis is known only for solutions of
type (\ref{Ell1})--(\ref{Ell11}) and solutions with nontrivial
phase of type (\ref{SSol1}) and (\ref{SSol2}). Linear analysis of
soliton solutions is well developed, but it is out scope of the
present paper.

Finally we discuss three special cases:

{\bf Case  I.} Let $B_0=B_j=0$ then for $j=1,\ldots, N_{f}$ and $
q_0=\sqrt{A_0}\sn(\alpha x,k)$,
 $q_j=\sqrt{A_j}\sn(\alpha x,k)$ we have the following
 linearized equations:
\begin{gather*}
\hbar\phi^{\rm{R}}_{0,t}=-\frac{1}{2m_{\rm{B}}}\partial^2_{xx}\phi^{\rm{I}}_0
+\left(V_0+g_{\rm{BB}}A_0+g_{\rm{BF}}\sum_jA_j\right)\sn^2(\alpha
x,k)\phi^{\rm{I}}_0
-\omega_0\phi^{\rm{I}}_0,\\
\hbar\phi^{\rm{I}}_{0,t}=\frac{1}{2m_{\rm{B}}}\partial^2_{xx}\phi^{\rm{R}}_0
-\left(V_0+3g_{\rm{BB}}A_0+g_{\rm{BF}}\sum_jA_j\right)\sn^2(\alpha
x,k)\phi^{\rm{R}}_0 \nonumber\\
\phantom{\hbar\phi^{\rm{I}}_{0,t}=}{}+\omega_0\phi^{\rm{R}}_0-2g_{\rm{BF}}A_0\sn^2(\alpha
x,k)\sum_j\phi^{\rm{R}}_j,
\\\hbar\phi^{\rm{R}}_{j,t}=-\frac{1}{2m_{\rm{F}}}\partial^2_{xx}\phi^{\rm{I}}_j
+ \left(V_0+g_{\rm{BF}}A_0\right)\sn^2(\alpha x,k)\phi^{\rm{I}}_j
-\omega_j\phi^{\rm{I}}_j,\\
\hbar\phi^{\rm{I}}_{j,t}=\frac{1}{2m_{\rm{F}}}\partial^2_{xx}\phi^{\rm{R}}_j
-\left(V_0+g_{\rm{BF}}A_0\right)\sn^2(\alpha x,k)\phi^{\rm{R}}_j
+\omega_j\phi^{\rm{R}}-2g_{\rm{BF}}\sqrt{A_0A_j}\sn^2(\alpha
x,k)\phi^{\rm{R}}_0.
\end{gather*}

{\bf Case II.} Let $B_0=-A_0$, $B_j=-A_j$ then for $
q_0=\sqrt{-A_0}\cn(\alpha x,k)$, $ q_j=\sqrt{-A_j}\cn(\alpha x,k)$
we obtain the following linearized equations:
\begin{gather*}
\hbar\phi^{\rm{R}}_{0,t}=-\frac{1}{2m_{\rm{B}}}\partial^2_{xx}\phi^{\rm{I}}_0
+\left(V_0+g_{\rm{BB}}A_0+g_{\rm{BF}}\sum_jA_j\right)\sn^2(\alpha
x,k)\phi^{\rm{I}}_0\nonumber\\
\phantom{\hbar\phi^{\rm{R}}_{0,t}=}{}-\left(g_{\rm{BB}}A_0+g_{\rm{BF}}\sum_jA_j+\omega_0\right)\phi^{\rm{I}}_0,\\
\hbar\phi^{\rm{I}}_{0,t}=\frac{1}{2m_{\rm{B}}}\partial^2_{xx}\phi^{\rm{R}}_0
+\left(3g_{\rm{BB}}A_0+g_{\rm{BF}}\sum_jA_j+\omega_0\right)\phi^{\rm{R}}_0
\nonumber\\
\phantom{\hbar\phi^{\rm{I}}_{0,t}=}{} -\left(V_0+3g_{\rm{BB}}A_0+g_{\rm{BF}}\sum_jA_j\right)\sn^2(\alpha
x,k)\phi^{\rm{R}}_0
+2g_{\rm{BF}}A_0\big(1-\sn^2(\alpha x,k)\big)\sum_j\phi^{\rm{R}}_j,\\
\hbar\phi^{\rm{R}}_{j,t}=-\frac{1}{2m_{\rm{F}}}\partial^2_{xx}\phi^{\rm{I}}_j
 + (V_0+g_{\rm{BF}}A_0)\sn^2(\alpha x,k)\phi^{\rm{I}}_j-(g_{\rm{BF}}A_0+\omega_j)\phi^{\rm{I}}_j,\\
\hbar\phi^{\rm{I}}_{j,t}=\frac{1}{2m_{\rm{F}}}\partial^2_{xx}\phi^{\rm{R}}_j
-(V_0+g_{\rm{BF}}A_0)\sn^2(\alpha x,k)\phi^{\rm{R}}_j+(g_{\rm{BF}}A_0+\omega_j)\phi^{\rm{R}}_j\nonumber\\
\phantom{\hbar\phi^{\rm{I}}_{j,t}=}{} -2g_{\rm{BF}}\sqrt{A_0A_j}\big(1-\sn^2(\alpha
x,k)\big)\phi^{\rm{R}}_0,\qquad j=1,\ldots, N_{f}.
\end{gather*}

{\bf Case III.} Let $B_0=-A_0/k^2$, $B_j=-A_j/k^2$ therefore the
solutions are
\begin{gather*}
q_0=\sqrt{-A_0}\dn(\alpha x,k)/k,\qquad q_j=\sqrt{-A_j}\dn(\alpha
x,k)/k,
\end{gather*}
and we obtain the following linearized equations
\begin{gather*}
\hbar\phi^{\rm{R}}_{0,t}=-\frac{1}{2m_{\rm{B}}}\partial^2_{xx}\phi^{\rm{I}}_0
+\left(V_0+g_{\rm{BB}}A_0+g_{\rm{BF}}\sum_jA_j\right)\sn^2(\alpha
x,k)\phi^{\rm{I}}_0\nonumber\\
\phantom{\hbar\phi^{\rm{R}}_{0,t}=}{} -\left(g_{\rm{BB}}A_0+g_{\rm{BF}}\sum_jA_j+k^2\omega_0\right)
\frac{\phi^{\rm{I}}_0}{k^2},\\
\hbar\phi^{\rm{I}}_{0,t}=\frac{1}{2m_{\rm{B}}}\partial^2_{xx}\phi^{\rm{R}}_0
+\left(3g_{\rm{BB}}A_0+g_{\rm{BF}}\sum_jA_j+k^2\omega_0\right)\frac{\phi^{\rm{R}}_0}{k^2},
\nonumber\\
\phantom{\hbar\phi^{\rm{I}}_{0,t}=}{} -\left(V_0+3g_{\rm{BB}}A_0+g_{\rm{BF}}\sum_jA_j\right)\sn^2(\alpha
x,k)\phi^{\rm{R}}_0
+\frac{2g_{\rm{BF}}A_0(1-k^2\sn^2(\alpha,k))}{k^2}\sum_j\phi^{\rm{R}}_j,\\
\hbar\phi^{\rm{R}}_{j,t}=-\frac{1}{2m_{\rm{F}}}\partial^2_{xx}\phi^{\rm{I}}_j
 + \left(V_0+g_{\rm{BF}}A_0\right)\sn^2(\alpha x,k)\phi^{\rm{I}}_j
 -\frac{g_{\rm{BF}}A_0+k^2\omega_j}{k^2}\phi^{\rm{I}}_j,\\
\hbar\phi^{\rm{I}}_{j,t}=\frac{1}{2m_{\rm{F}}}\partial^2_{xx}\phi^{\rm{R}}_j
-\left(V_0+g_{\rm{BF}}A_0\right)\sn^2(\alpha x,k)\phi^{\rm{R}}_j
+\frac{g_{\rm{BF}}A_0+k^2\omega_j}{k^2}\phi^{\rm{R}}_j \nonumber\\
\phantom{\hbar\phi^{\rm{I}}_{j,t}=}{} -\frac{2g_{\rm{BF}}\sqrt{A_0A_j}\big(1-k^2\sn^2(\alpha,k)\big)\phi^{\rm{R}}_0}{k^2},\qquad
j=1,\ldots, N_{f}.
\end{gather*}
These cases are by no means exhaustive.

\section{Conclusions}\label{sec:8}
In conclusion, we have considered the mean f\/ield model for
boson-fermion mixtures in optical lattice. Classes of
quasi-periodic, periodic, elliptic solutions, and solitons have
been analyzed in detail. These solutions can be used as initial
states which can generate localized matter waves (solitons)
through the modulational instability mechanism. This important
problem is under consideration.

\subsection*{Acknowledgements} The present work is supported by the
National Science Foundation of Bulgaria,  contract No F-1410.

\pdfbookmark[1]{References}{ref}
\LastPageEnding

\end{document}